\documentclass[
 reprint,
aps,
prc,
floatfix]{revtex4-1}

\usepackage{graphicx}
\usepackage{dcolumn}
\usepackage{bm}

\date { }

\begin{document}

\preprint{APS/123-QED}

\title{Ground state properties of $^4$He and $^{12}$C nuclei at equilibrium and at large static compression at zero temperature using Nijmegen and Reid Soft Core nucleon-nucleon interactions
}

\author{Iyad Alhagaish$^{1}$, Ali Abu-Nada$^{2}$,  Feras Afaneh$^{1}$, and Mahmoud Hasan$^{3}$
}

\affiliation{
 $^{1}$Department of Physics, The Hashemite University, Zarqa, Jordan\\
 $^{2}$Department of Physics, Isra University, Amman, Jordan\\
    $^{3}$Aqaba Technical University
            }

\date{\today}

\begin{abstract}
In this paper, we investigate the ground state properties ($i.e.$ binding energy, nuclear radius, radial density distribution and single particle energies) for $^4$He and $^{12}$C nuclei at equilibrium and at large static compression at zero temperature by using two realistic different potentials namely, Nijmegen and Reid Soft Core (RSC)potentials. We carry out the calculations in No-Core Shell Model  space consisting of six major oscillator shells  within the framework of the Constrained Spherical Hartree-Fock (CSHF) approximations. We find out that, the computed equilibrium root mean square radii and the Hartree Fock energies for $^4$He and $^{12}$C  with those two different potentials are very close to the experimental values of the nuclear radii and nuclear binding energies for the same nuclei.
\end{abstract}

\maketitle


\section{\label{sec:level1}Introdcution}

The nucleus is composed of fermions of spin $1/2$ and isospin $1/2$, called nucleons (protons and neutrons), interacting with each other through a complex interaction
with a short-range repulsive core \cite{Bohr}. For the purpose of 
investigating nuclear structure, it is convenient to represent
the inter-nucleon interaction by a potential. Various
potential models for the two nucleons interaction has
been proposed \cite{Hasan,Vary,Faessler} and parameters in the model fitted
using the result of N-N scattering experiments. The central
problem of nuclear structure theory is the solution
of Many Body Schrödinger Equation (MBSE). For Hamiltonians
of interest in the nuclear case, an analytic solution
is impossible (except in certain simple cases) and
one is compelled to use some approximation, either in
the numerical solution of the equation or specification of
the Hamiltonian, or both \cite{Stone}. One has to solve the full
(MBSE) numerically in an exact way as possible using
variational techniques. No-Core Shell Model (NCSM)
calculations have been made in light nuclei, and heavier
nuclei close to closed shell have been treated \cite{Bhaduri}.

The NCSM is based on a new variation of the well known
shell model for nuclei. Historically, shell-model
calculations have been made assuming closed, inert core
of nucleons with only a few active valence nucleons. The
interaction of these valence nucleons with the core and
with other valence nucleons could not be described by
microscopic interactions, as they have been developed for
few-nucleon systems. Therefore, these attempts have not
been completely successful in relating the effective shell model
interaction to the basic nuclear interaction. This
situation has been changed in 1990 with the development of
the NCSM, which treats all nucleons in the nucleus as an
active particles. The NCSM assumes that all nucleons
are active, there is a systematic way to obtain the effective
6interactions from bare NN and 3N forces. This is
the strength of the NCSM compared to traditional shell model calculation \cite{Wiring}.

The ab-initio No-Core NCSM has been applied with realistic
effective N-N interactions to light nuclei \cite{Barrett96,Barrett00}. It
has been shown that the NCSM approach can be consistently
applied to solve the three-nucleon as well as
four-nucleon bound state problem. There are various
models for nuclear potentials such as Bonn, CD-Bonn,
Paris, Nijmegen, and Idaho. They all describe the observed
deuteron and N-N scattering data very accurately.
However, due to their strong repulsive core, none of them
can be used directly in the nuclear structure calculations.
To overcome this difficulty, the Brueckner G-matrix has
been used traditionally as a starting point, but as is well
known its energy dependence is an undesirable feature,
in particular dealing with Hartree-Fock calculations \cite{Kuo}.

Hartree-Fock is a proven tool for semi-realistic interactions
even for the heaviest nuclei \cite{Schuck} and is sufficiently
flexible to handle many-body forces \cite{Vary94,Vary97}. Also, It is
a starting point for practical many body methods used
extensively in heavier systems \cite{Schuck}. Of course , there is
a long history going back to Brueckner of merging the
mean field method with non relativistic effective potentials
(G-matrix) derived from N-N interactions \cite{McCarthy,Day}.

We investigate the ground state properties of $^{4}$He and $^{12}$C nuclei at equilibrium and at large amplitude of compression with zero temperature using a realistic effective interaction based on two different potentials namely, Nijmegen and Reid Soft Core
(RSC) potentials. We perform the calculations in NCSM  space consisting of six major
oscillator shells ($i.e.$ 21 single particles orbitals) within the framework of the constrained spherical Hartree-Fock (CSHF) approximations. In particular, we study
the sensitivity of the ground state properties; such as
binding energy, nuclear radius, radial density distribution
and single particle energies to the degree of compression.
The importance of this study is to investigate the effect
of the potential used on softening the nuclear equation of state. 

This study also will shed some light on the behavior of nuclear matter under extreme conditions, which has its importance in astrophysical problems and
to have better understanding of its behavior in nucleus-nucleus collisions as in heavy ion collisions in high-energy super-colliders. On the other hand, there are important
physical motivations for investigating $^{12}$C. Actually, $^{12}$C
nucleus plays an important role in neutrino studies as it is an ingredient of neutrino liquid-scintillator detectors \cite{Barrett00}.

\section{\label{sec:level2}Methodlogy and parameters}
Our  nuclear system consists of $A$  Nucleons ($N$ neutrons
and $Z$ protons) of spin $s = 1/2$, isospin $\tau = 1/2$
and mass $m$ each. The Hamiltonian of the system consists
of the single particle energy and a two- body interaction:
\begin{equation} \label{eq:1}
H= \sum^{A}_{i=1} t_{i} + \sum^{A} _{i<j} V_{ij}
\end{equation}
Where $t$ denotes the single particle kinetic energy operator, and $V_{ij}$ is the two-body interaction term which consists of two body interaction and the Coulomb potential
$\left(V = V_{NN} + V_{C} \right)$. The labels enclosed within the
brackets  refer to particle coordinates, where the restriction
$i < j$ in the second sum in equation (\ref{eq:1}) take
care of the fact that the interaction has to be summed
counting each pair only once. In this work, we use a no core-effective
Hamiltonian; that is all nucleons are active.

In principle, if one solves the many-body problem in
the full Hilbert space, then one gets the exact solution, however,  this is not possible for nuclei with $A > 4$. Therefore, we truncate the Hilbert space to finite model space. The
price one has to pay is to define an effective Hamiltonian $ \left( H_{eff} \right)$ based on the study presented in \cite{Vary84, Vary85}.

	The detailed calculations of the  effective Hamiltonian, $H_{eff}$, model space,  calculation procedures, and strategy  have been extracted   from  references \cite{Vary84, Vary85}. Based on these studies, the two-body matrix elements are scaled to an
optimal value of $\omega$, the adjusting parameters $ (\lambda_{1}, \lambda_{2},$
and $\hbar  \omega \prime)$ for $^{4}He$ and $^{12}C$ nuclei in a given model space
at equilibrium, are presented  in Table \ref{table:1}. We notice that
the value of $\lambda_{1}$  is less than unity, this is because the fact that 
kinetic energy operator ($T_{rel}$) is a positive definite operator
and if it is normalized by itself into a finite model
space this will reduce its magnitude. Moreover, we notice
that the value of $\lambda_{2}$ is greater than unity in order
to compensate for the lack of sufficient binding when we
truncate the full Hilbert space to a finite model space. In
our calculations, we use a large model space consisting
six major shells; $i.e.$  21 nucleon orbitals each orbital has
definite quantum numbers, $n, ℓ, s, J$.
\begin{table}
\centering
\caption{Values of adjusting Parameters of the effective
Hamiltonian ($H_{eff}$) for $^{4}He$ and $^{12}C$ by using Nijmegen and RSC penitentials in six shells to get an agreement between HF results and experimental data \cite{Hasan87,Hale92}.
}

\begin{tabular}{c c c c c} 
 \hline \hline 
Nucleus & Potential & $\hbar  \omega \prime$ & $\lambda_{1}$ &$\lambda_{2}$  \\ [0.5ex] 
 \hline
$^{4}He$  & Nijmegen  &  15.700& 0 0.990 & 1.041 \\ 
 &  RSC & 17.872&0.980& 1.186\\ 
 $^{12}C$ &  Nijmegen & 8.454 &0.976 &1.200 \\
  &  RSC&   10.104& 0.973& 1.420 \\ [1ex] 
  
 \hline \hline
\end{tabular}
\label{table:1}
\end{table}
\section{\label{sec:level3} results and discussion}
\subsection{\label{subsec:level1} Results of $^{4}He$}

The parameters $\lambda_{1}$, $\lambda_{2}$, and $\hbar  \omega \prime$ which have been
used for $^{4}He$ are presented  in Table \ref{table:1}. With these parameters
we find  that the equilibrium root mean square radius $(r_{rms}) $ and $E_{HF}$ using RSC (Nijmegen) potentials are, $r_{rms}$ = 1.46fm ($r_{rms}$ = 1.46fm), and $E_{HF}$ = -28.296MeV ($E_{HF}$ = -28.296MeV),  respectively. We remind the reader here that the experimental nuclear radius for $^{4}He$, is $r_{rms}$ = 1.46fm and the measured binding energy is $E_{bind}$ = -28.296MeV [18, 19]. We found
that the occupied orbitals are $0s_{1/2}$ in agreement with the standard shell model.

In Figure 1, the $E_{HF}$ energies  using RSC and Nijmegen potentials are presented  as a function of  $r_{rms}$. In Figure 2, the single particle energies $SPEs$  are displayed as a function of   $r_{rms}$. Moreover, Figure 3 represents   the radial density distribution for neutrons ($\rho_{n}$), protons ($\rho_{p}$), and $\rho_{total} = \rho_{n} + \rho_{p}$, at equilibrium ($i.e.$ at $r_{rms} = 1.46fm$) using Nijmegen potential, while Figure 4 displays the total radial density distribution $\rho_{total}$ at equilibrium ($r_{rms} = 1.46fm$) and at large static compression ($r_{rms} = 1.33fm$) using Nijmegen potential. In addition, Figure 5 displays the total radial density distribution at equilibrium ($r_{rms} = 1.46fm$) and at large static compression ($r_{rms} = 1.24fm$) using RSC potential. In Figure 6, we compare $\rho_{total}$ for the two potentials (Nijmegen and RSC) at equilibrium ($r_{rms} = 1.46fm$). Finally, In Figure 7 we compare $\rho_{total}$ for the two potentials (Nijmegen and RSC) at large static compression ($r_{rms} = 1.33fm$).

\begin{figure}
\caption{ $E_{HF}$   using RSC and Nijmegen potentials in six-oscillator Shells as a function of  $r_{rms}$ for $^{4}He$}
\centering
\includegraphics[width=0.5\textwidth]{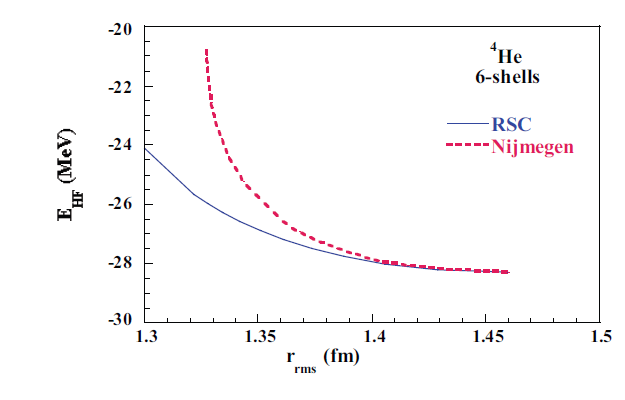}
\end{figure}

\begin{figure}
\caption{    Single particle energy (S.P.E) for $^{4}He$ in six-oscillator shells as a fucntion of  $r_{rms.}$ using Nijmegen and (RSC) potentials.}
\centering
\includegraphics[width=0.5\textwidth]{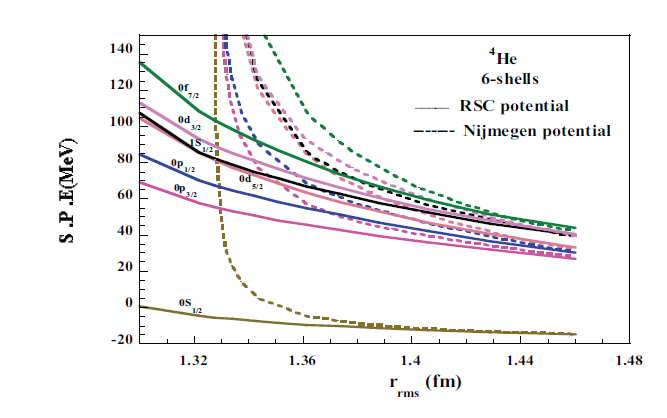}
\end{figure}

\begin{figure}
\caption{Radial density distribution for $4^{He}$
 as a function of nucleus  radius r(fm) at equilibrium (r = 1.46fm). Using Nijmegen potential.
0}
\centering
\includegraphics[width=0.5\textwidth]{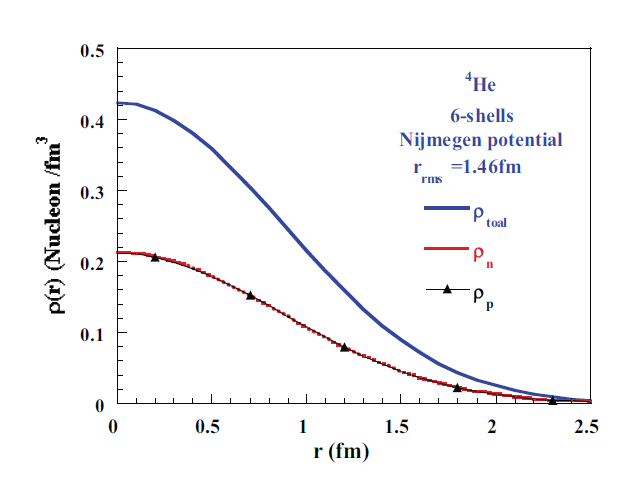}
\end{figure}

\begin{figure}
\caption{Total radial density distribution $\rho_{total}$
for $^{4}He$ at equilibrium ($r_{rms} = 1.46fm$) and large static compression ($r_{rms} = 1.33fm$) as a function of nucleus radius r(fm). Using Nijmegen potential.}
\centering
\includegraphics[width=0.5\textwidth]{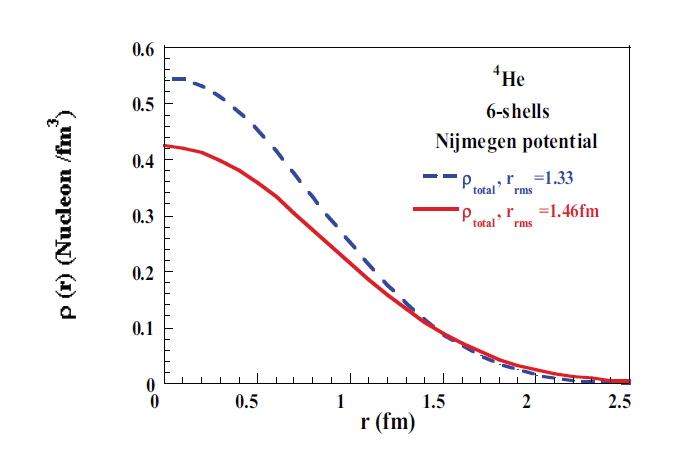}
\end{figure}

\begin{figure}
\caption{Total radial density distribution $\rho_{total}$
for $^{4}$He at equilibrium ($r_{rms} = 1.46fm$) and large static compression ($r_{rms} = 1.24fm$) as a function r(fm) using RSC potential.}
\centering
\includegraphics[width=0.5\textwidth]{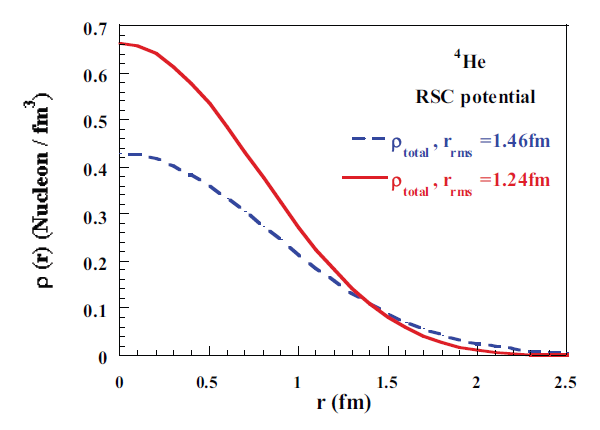}
\end{figure}

\begin{figure}
\caption{Total radial density distribution $\rho_{total}$
for $^{4}He$ at equilibrium ($r_{rms} = 1.46fm$) using Nijmegen and RSC potentials.}
\centering
\includegraphics[width=0.5\textwidth]{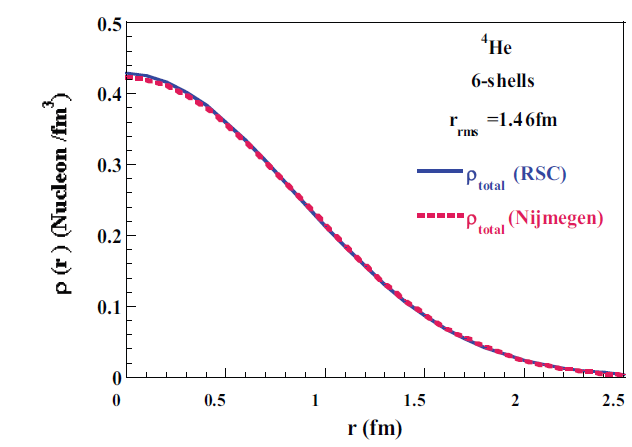}
\end{figure}

\begin{figure}
\caption{Total radial density distribution $\rho_{total}$
for $^{4}He$ at large compression ($r_{rms} = 1.33fm$) using Nijmegen and RSC potentials.}
\centering
\includegraphics[width=0.5\textwidth]{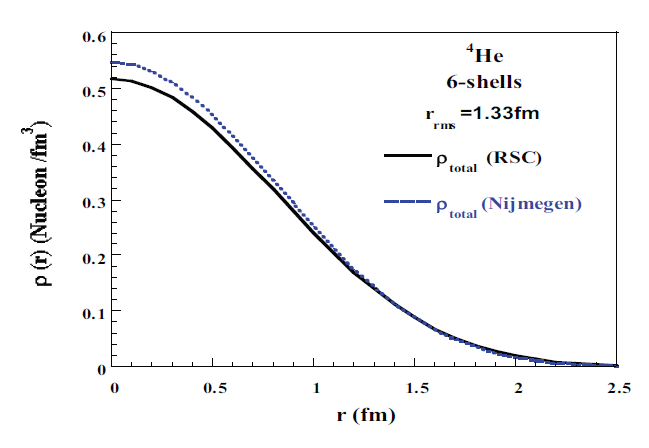}
\end{figure}

In Figure 1, we notice that, using RSC potential that
reducing the volume of the nucleus by about $24\%$ (at
equilibrium $r_{rms} = 1.46fm$, and at large static compression, $r_{rms} = 1.33fm$), compared to its volume at
equilibrium reduces the binding energy by about $8\%$ (at
equilibrium, $E_{HF} = -28.296MeV$, and at large static
compression $E_{HF} = -25.956MeV$), but when the nuclear
volume is reduced by about $417\%$, the change in the
nuclear binding energy by about $3\%$. However, when we
use Nijmegen potential that will reduce the volume by
about $24\%$ compared to its volume at equilibrium reduces
the binding energy by about $25\%$, and when we reduce
the nuclear volume by about $17\%$ the change in the nuclear
binding energy by about $4\%$. This means that the nuclear equation of state becomes stiffer as we compress
the nucleus. We also note that, at large compression the
nuclear binding energy using Nijmegen potential is larger
than the binding energy using RSC potential. In Figure 2, we notice that, the ordering of the orbits is in exact agreement with the orbits ordering of the standard
shell model. The gap is very clear between the shells, and also the splitting of the levels in each shell is an indicator that the L-S coupling is strong enough in RSC and Nijmegen potentials. This is also clear from shifting down the $0f_{7/5}$ orbit from the p-f shell to the s−d shell. The L-S coupling becomes stronger as we increase the static load
on the nucleus. Finally, we notice that, the levels curve up
as we compress the nucleus more and more and the levels
curve up more rapidly using Nijmegen potential. This
means that the kinetic energy becomes more influential
than the attractive means field of the nucleon.
Figure 3, shows the radial density distribution for neutrons
$\rho{_n}$ , protons $\rho_{p}$ ,and their sum $\rho_{total}$ as a function of the nucleus radius $r$ at equilibrium ($i.e.$ $r_{rms} = 1.46fm$). We conclude that, the neutron and proton densities are almost the same except in the interior region, where the neutrons are denser than protons. Actually, this difference in densities is attributed to the Coulomb repulsion between the protons.
Figure 4,  represents the total nuclear radial density
distribution $\rho_{total}$ at equilibrium ($i.e.$ $r_{rms} = 1.46fm$) and at ($r_{rms} = 1.33fm$) (a volume reduction by about 7$5\%$) using Nijmegen potential. The nuclear density becomes denser in the interior and less dense in the exterior, ($i.e.$ close to the surface of the nucleus). This means that,  as we increase the load more and more; the
surface of the nucleus becomes more and more responsive.
Figure 5,  displays the total nuclear radial density
distribution $\rho_{total}$ at equilibrium ($i.e.$ $r_{rms} = 1.46fm$) and at ($r_{rms} = 1.24fm$) (a volume reduction by about $61\%$) using RSC potential. In addition, Figure 6 displays the total radial density distribution total at equilibrium ($r_{rms} = 1.46fm$), at two different potentials (Nijmegen and RSC). We notice that, in the interior region the $\rho_{total}$ using RSC potential is larger than $\rho_{total}$ using Nijmegen potential, but this difference is very small. In the exterior region $\rho_{total}$ for both potential are nearly the same.
 Figure 7, displays the total radial density
distribution $\rho_{total}$  at large static compression ($r_{rms} = 1.33fm$) for the two different potentials (Nijmegen and RSC). Finally, We  see, first, it is clear that from Figures 6 and 7 as the nucleus is compressed; the
nuclear density become denser in the interior and less
dense in the exterior for both potentials (Nijmegen and
RSC), and second, the nuclear density becomes denser in the
interior with Nijmegen potential than with RSC potential,
and less dense in the exterior with Nijmegen potential
than with RSC potential. This means that as
we increase the load more and more, the surface of the
nucleus becomes more and more responsive, and it is possible
to compress the nucleus to a smaller radius using
RSC potential than Nijmegen potential.
\subsection{\label{subsec:level2} Results of $^{12}C$}

The calculations proceed in the same manner as the
calculations for $^4{He}$. The values of the parameters $\lambda_{1}$,
$\lambda_{2}$, and $\hbar \prime{\omega}$ to obtain the agreement between the $E_{HF}$ and $r_{rms}$, and the experimental binding energy and the
nuclear radii \cite{Ajzenberg} are listed in Table \ref{table:1}. In the input
data file, we change the mass number from $4$ to $12$, number of protons and neutrons are $6$, and the occupied orbits are $0s_{1/2}$ and $0p_{3/2}$ . With the adjusting parameters ($i.e.$  $\lambda_{1}$, $\lambda_{2}$, and $\hbar \prime{\omega}$),  we find an equilibrium ($r_{rms}$) and $E_{HF}$ using RSC (Nijmegen) potentials, $r_{rms} = 2.3508fm$ ($r_{rms} = 2.3498fm$), and $E_{HF} =
-92.174MeV$ ($E_{HF} = −92.167$), respectively. The experimental
nuclear radius $r_{exp} = 2.35fm$ and the binding
energy is $E_{bind} = -92.162MeV$ \cite{Ajzenberg}. We find the
occupied orbital are $0s_{1/2}$ and $0p_{3/2}$, in agreement with the standard shell model.

The $E_{HF}$ energies as a function of  $r_{rms}$ using RSC and Nijmegen
potential are displayed in Figure 8. The single particle energies (SPE) as a function of  the $r_{rms}$ are displayed in Figure 9. Moreover, Figure 10 displays the radial density distribution for neutrons $\rho_{n}$, protons $\rho_{p}$, and $\rho_{total} = \rho_{n} + \rho_{p}$, at equilibrium ($i.e.$ at $r_{rms} = 2.35fm$) using Nijmegen potential. Figure 11 represents the radial
density distribution at equilibrium($r_{rms} = 2.35fm$) and at large static compression ($r_{rms} = 2.26fm$) using Nijmegen potential. In Figure 12, we compare $\rho_{total}$ at two different values of $r_{rms}$; at $r_{rms} = 2.35fm$ ($i.e.$ equilibrium) and at $r_{rms} = 2.06fm$ using RSC potential. In
Figure 13, we compare $\rho_{total}$ at two different potentials (Nijmegen and RSC) at equilibrium ($r_{rms} = 2.35fm$).
Furthermore, in Figure 14, we compare total at two different
potentials (Nijmegen and RSC) at large static compression
($r_{rms} = 2.26fm$). 

We notice in Figure 8, that, reducing the volume of the nucleus to about $12\%$ compared to its volume at equilibrium reduces the binding
energy by about $2\%$ using RSC potential. However, when reduce the nuclear volume by about $6\%$ the change in the nuclear binding energy by about $1\%$. But, by using Nijmegen potential for the same reduction in nuclear volume compared to the volume at equilibrium reduces the binding energy by $11\%$ and $1\%$. Thus, that means that the nuclear equation of state becomes stiffer as we
compress the nucleus, using RSC potential softening the
equation of state more than using Nijmegen potential. In Figure 9, we notice that the ordering of the orbits in agreement with the orbital ordering of the standard
shell model. In addition, we notice that the splitting of
the levels in each shell is an indicator that the L-S
coupling is strong enough in both potentials. This is also
clear from shifting down the $0f_{7/5}$ orbit from the p-f
shell to the s-d shell. This L-S coupling becomes
weaker as we increase the static load on the nucleus. On the other hand, the orbits curve up as we increase the load on the nucleus. The SPE levels curve up more rapidly when using Nijmegen potential. This was discussed
in the results of $^4{He}$, and that is because the
kinetic energy of the nucleon which is positive quantity
becomes more influential than the attractive mean field of the nucleon. In addition, we realize that the SPE's when using Nijmegen potential are less bound than the (SPE) with using RSC potential, especially when we compress
the nucleus more and more. Figure 10 shows the radial density distribution for neutron $\rho_{n}$ , protons $\rho_{p}$ ,and their sum $\rho_{total}$ as a function of the radial distance from the center of the nucleus at
equilibrium ($i.e.$ $r_{rms} = 2.35fm$). We notice that the
neutron's and proton's densities are almost the same except
in the interior region, where the neutrons are denser
than protons. This difference in densities is attributed
to Coulomb repulsion between the protons. Also, Figure
11 displays the total radial density distribution $\rho_{total}$
nucleus at equilibrium ($i.e.$ $r_{rms} = 2.35fm$) and at ($r_{rms} = 2.26fm$, the volume reduction is about $11\%$) using Nijmegen potential. Obviously, that as the nucleus is compressed; the nuclear density becomes denser in the
interior and less dense in the exterior ($i.e.$ closer to the
surface of the nucleus). This means that, as we increase
the load more and more, the surface of the nucleus becomes
more responsive.\\
Figure 12 displays the nuclear total radial density distribution
$ρ_{total}$ at equilibrium ($i.e.$ $r_{rms} = 2.35fm$) and
at ($r_{rms} = 2.06fm$, the volume reduction is about $67\%$)
using RSC potential. Interestingly, we notice the same
features as in Figure 11, that is the interior of the nucleus
becomes more dense than the exterior as we compress the
nucleus. There is one difference though; that the increase
in nuclear density is more pronounced using RSC than when using Nijmegen potential. \\
Figure 14 displays the total radial density distribution
$ρ_{total}$ at large static compression ($r_{rms} = 2.26fm$) with
the two potentials (Nijmegen and RSC).We draw the following
conclusions. First, it is clear that as the nucleus
compressed; the nuclear density become denser in the interior
and less dense in the exterior for both potentials.
Second, the increase in nuclear density with compression
in the interior is more pronounced using RSC potential
than with Nijmegen potential and less dense in the exterior
using RSC potential than with Nijmegen potential.
Finally, Figure 13 displays the total radial density distribution
$\rho_{total}$ at equilibrium ($r_{rms} = 2.35fm$) with the
two potentials (Nijmegen and RSC). We notice from this
figure that the interior region the nuclear density is larger
using RSC than when using Nijmegen potential. The situation
is reversed in the exterior.

\begin{figure}
\caption{Constrained spherical Hartree-Fock
energy (CSHFE) For $^12{C}$ in six-oscillator Shells as a function of  $r_{rms}$.
Using RSC and Nijmegen potentials}
\centering
\includegraphics[width=0.5\textwidth]{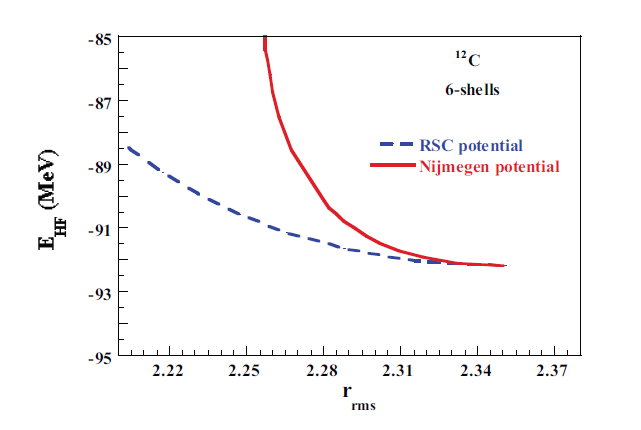}
\end{figure}

\begin{figure}
\caption{Single particle energy (S.P.E) for $^12{C}$
in six-oscillator shells as a function of $r_{rms}$ Using Nijmegen and (RSC)
potentials.}
\centering
\includegraphics[width=0.5\textwidth]{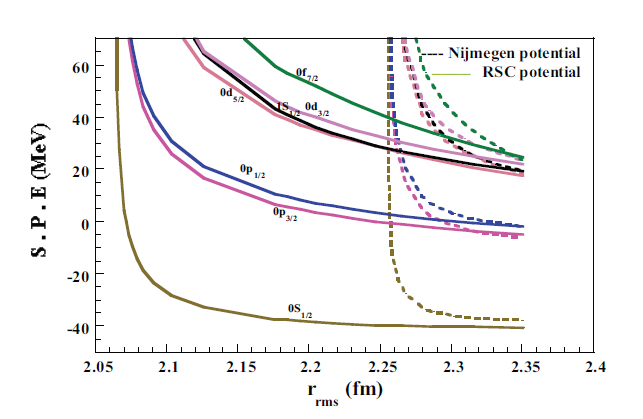}
\end{figure}

\begin{figure}
\caption{Radial density distribution for $^{12}{C}$
as a function of  nuclear radius r(fm) at equilibrium ($r_{rms} = 2.35fm$) using Nijmegen potential.}
\centering
\includegraphics[width=0.5\textwidth]{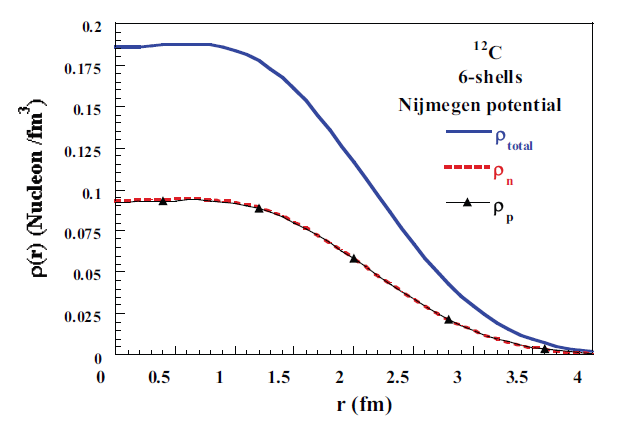}
\end{figure}

\begin{figure}
\caption{Total radial density distribution $\rho_{total}$
for $^{12}{C}$ at equilibrium ($r_{rms} = 2.35fm$) and large static compression
($r_{rms} = 2.26fm$) as a function of nuclear radius  r(fm). using Nijmegen potential.}
\centering
\includegraphics[width=0.5\textwidth]{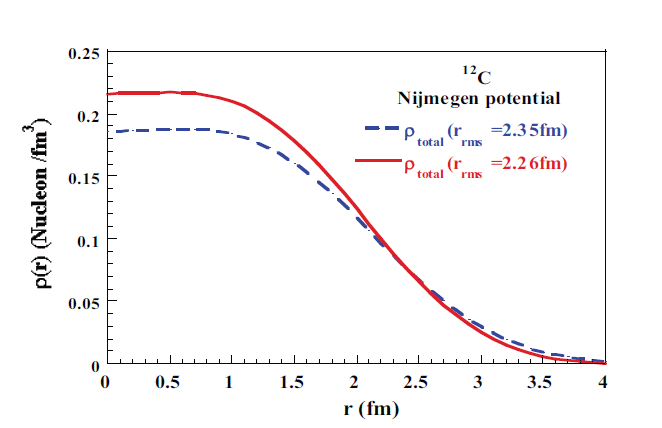}
\end{figure}

\begin{figure}
\caption{Total radial density distribution $\rho_{total}$
for $^{12}{C}$ at equilibrium ($r_{rms} = 2.35fm$) and at large static
compression ($r_{rms} = 2.06fm$) as a function of nuclear radius r(fm) using RSC potential.}
\centering
\includegraphics[width=0.5\textwidth]{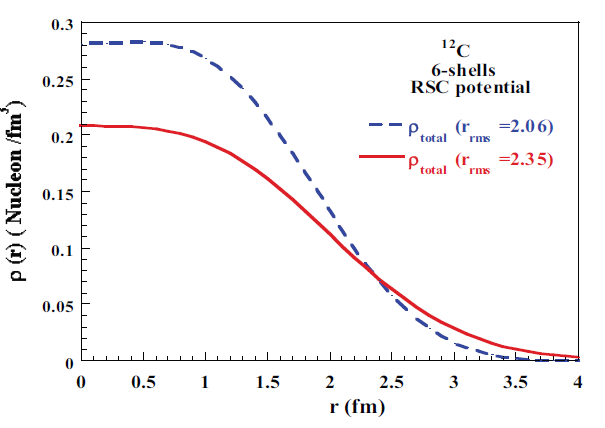}
\end{figure}

\begin{figure}
\caption{Total radial density distribution $\rho_{total}$
for $^{12}{C}$ at equilibrium ($r_{rms}  = 2.35fm$) as a function of nuclear radius r(fm) using Nijmegen and RSC potentials.}
\centering
\includegraphics[width=0.5\textwidth]{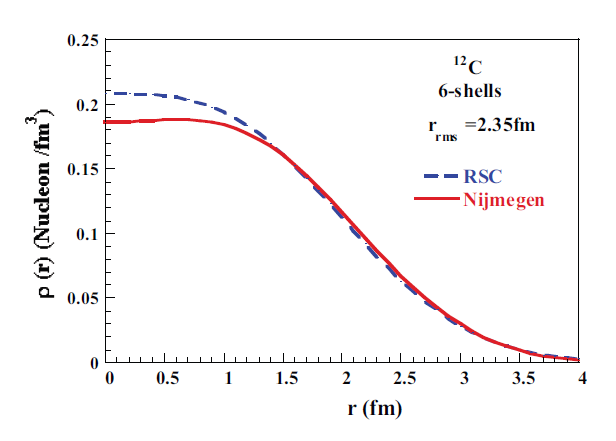}
\end{figure}

\begin{figure}
\caption{Total radial density distribution $\rho_{total}$
for $^{12}{C}$ at large static compression ($r_{rms} = 2.26fm$) using
Nijmegen and RSC potentials.}
\centering
\includegraphics[width=0.5\textwidth]{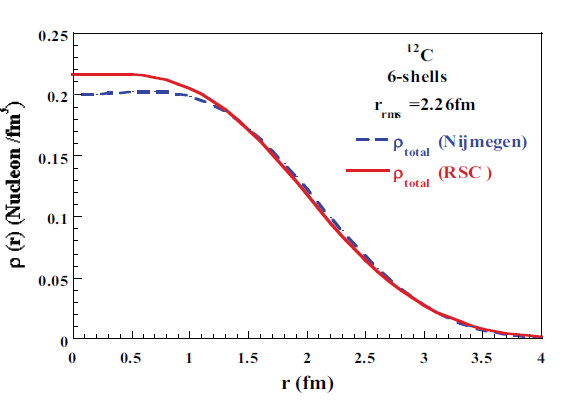}
\end{figure}

\section{\label{sec:level3}conclusion}

The ground state properties of $^{4}He$ and $^{12}C$ nuclei have been investigated at equilibrium and at large amplitude of compression using a realistic effective interaction based on two different potentials namely, Nijmegen and Reid Soft Core (RSC). The calculations were performed in no-Core model space consisting of six major oscillator shells ($i.e.$ 21 single particles orbitals) within the framework of the constrained spherical Hartree-Fock (CSHF) approximations. Specifically, the sensitivity of the ground state properties, such as binding energy, nuclear radius, radial density distribution and the single particle energy, to the degree of compression were investigated.\\
We find that, the equilibrium root mean square  radius $r_{rms}$
for $^{4}He$ equals to $1.46fm$, and the corresponding Hartee-Fock Energy $E_{HF}$ equals to $-28.296MeV$ using both potentials (Nijmegen
and RSC) in a good agreement with the experimental
results of nuclear radius of $r_{rms} = 1.46fm$, and experimental
binding energy of $-28.296MeV$ \cite{Hasan87,Hale92}. For
the case of $^{12}C$, we find that the equilibrium $rms$ radius
equals to $2.351fm$ and $2.35fm$, where the corresponding
$E_{HF}$ are $-92.174MeV$ and $-92.167MeV$ using RSC
and Nijmegen potentials, respectively. However, the experimental
nuclear radius for $^{12}C$ equals to $2.35fm$ and
value of the binding energy equals to $-92.162MeV$ \cite{Ajzenberg}.\\
For $^{4}He$, with maximum compression used the minimum
$r_{rms}$ radii obtained are $1.244fm$, and $1.327fm$ and the
corresponding EHF are $-11.260MeV$ and $-20.804MeV$
using RSC and Nijmegen potentials respectively. On the
other hand, for $^{12}C$, the minimum $r_{rms}$ radii obtained
are $2.063fm$, and $2.255fm$, and the corresponding $E_{HF}$
are $-49.579MeV$ and $-82.444MeV$ for using RSC and
Nijmegen potentials respectively. For both nuclei, it is possible to compress the nucleus to a smaller size using
RSC than using Nijmegen potential. At equilibrium, the
neutrons and protons densities are almost the same except
in the interior region; where neutrons are more dense
than protons. This difference in densities is attributed to
Coulomb repulsion between protons. For $^{4}He$, and at
equilibrium, the radial density distribution are the same
except in the interior region using RSC and Nijmegen potential;
where it is larger with RSC than with Nijmegen
potential. At large compression the situation is reversed
especially in the interior region; the radial density distribution
becomes larger than the radial density distribution
when using RSC potential. For the case of $^{12}C$,
and at equilibrium, with the two potentials (Nijmegen
and RSC) the nuclear density is larger using RSC than
when using Nijmegen potential. The situation is reversed
in the exterior. Moreover, for both nuclei ($i.e.$ $^{4}He$ and
$^{12}C$), the ordering of the orbit is in exact agreement with
the orbit ordering of the standard shell model. The gap is
very clear between the shells. The splitting of the levels
in each shell is an indicator that L-S coupling is strong
enough in RSC and Nijmegen potentials. This is also
clear from shifting down the $0f_{7/5}$ orbit from the p-f
shell to the s-d shell. This L-S coupling becomes
is stronger as we increase the static load on the nucleus.
In addition, these levels curve up as we compress the
nucleus more and more and these levels curve up more
rapidly when using Nijmegen potential. This indicates
that the kinetic energy of the nucleus which is a positive
quantity becomes more and more influential than
the attractive means field of the nucleon. Finally, we notice
that the SPE’s becomes larger when using Nijmegen
potential than the (SPE) with using RSC potential, especially
at large compressions.

\end{document}